\documentclass[12pt]{article}
\begin{document}
\begin{center}
Saha Equation in Rindler Space

\bigskip
Sanchari De$^a$ and 
Somenath Chakrabarty$^{a\dagger}$

\medskip
$^a$Department of Physics, Visva-Bharati, Santiniketan-731235,
India \\
$^\dagger$E-mail: somenath.chakrabarty@visva-bharati.ac.in
\end{center}

\medskip
\begin{center}
Abstract
\end{center}

\medskip
\noindent The Saha equations for photo-ionization process of hydrogen atoms and the
creation of electron-positron pairs at high temperature are investigated
in a reference frame undergoing a uniform accelerated motion in an
otherwise flat Minkowski space-time geometry or equivalently, in
a rest frame in presence of a strong uniform gravitational field.
It is known as the Rindler space.


\section{Introduction}
The well known Lorentz transformations are
the space-time coordinate transformation between two inertial
frame of references 
\cite{LL}.
However, with the help of the principle of equivalence, one may obtain the
space-time transformations between a uniformly accelerated frame 
and an inertial frame and vice-versa
in the same manner as it is done in special theory
of relativity
\cite{WB,MTW,RL,MO}. In the present scenario the flat geometry is
called the Rindler space.  
For an illustration, let us consider  two reference frames- 
the frame $S$ is an inertial one,  which is at rest but
there is a uniform gravitational field and the frame $S^\prime$,
which is undergoing a uniform accelerated motion with respect to some inertial
frame, but in absence of gravitational field. Then 
in accordance with the principle of equivalence 
these two frames
are physically equivalent. Alternatively, one may state, that a
reference frame in uniform acceleration in absence of gravitational
field is equivalent to a frame at rest in presence of a uniform
gravitational field. We may assume that the strong gravitational field is produced by a black hole like gravitating
object. Therefore the gravitational field may be approximated by a
constant value within a small region at a distance $x$ from the centre
of the gravitating object. This is also called the local acceleration
of the frame.

To investigate Saha equation in a uniformly accelerated frame of
reference or in Rindler space,
we first develop a formalism 
with the physical concepts of principle of equivalence
to 
obtain the elements of the metric tensor
$g^{\mu\nu}$. 
We shall show that
analogous to the Minkowski space-time metric tensor, the off-diagonal 
elements of $g^{\mu\nu}$ are also zero in Rindler space.
In the next step, with the 
conventional form of action as has been defined in special
theory of relativity \cite{LL}, which is invariant here also,
the Lagrangian of a particle (in
our study, which may 
be a hydrogen atom or a hydrogen ion or an electron) is derived from
Hamilton's principle. Which
further gives the momentum and energy or Hamiltonian of the particle from the
standard relations of classical mechanics. Then considering a partially 
ionized 
hydrogen plasma which is a reactive mixture of neutral hydrogen atoms, 
hydrogen ions, electrons and
photons or considering an electron-positron plasma composed of an 
interacting  mixture
of electrons, positrons and photons, we shall obtain the modified
form of Saha equations for both the cases when observed from  a uniformly 
accelerated
frame of reference or in Rindler space. 
To the best of our knowledge, the study of Saha
equation in Rindler space  has not been reported earlier.
We shall also compare our findings with the conventional results.

The article is organized in the following manner: In the next section
we shall establish some of the useful relations of special theory of
relativity in uniformly accelerated frame. In this section we shall
also obtain the relativistic version of Hamiltonian in Rindler
coordinate system.
In section-3 we shall investigate photo-ionization of hydrogen atoms
in partially ionized hydrogen plasma in Rindler space.
In section-4 we shall study 
the electron-positron
pair production from intense electromagnetic radiation at high temperature
in an interacting electron-positron plasma, observed from the same
kind of uniformly accelerated frame.
In the last section we give the conclusion of our findings.
\section{Basic Formalism}
For the sake of completeness,
in this section, following \cite{MS,MAX,DP} we shall establish some of the useful formulas of
special theory of relativity for a uniformly accelerated frame of
reference.
Let us first assume that the frame $S^\prime$  has rectilinear motion with
uniform velocity $v$ along
$x$-direction with respect to some inertial frame $S$. Further the 
coordinates
of an event occurred at the point $P$ (say) is indicated by
$(x,y,z,t)$ in $S$-frame and with
$(x^\prime,y^\prime,z^\prime,t^\prime)$ in the frame $S^\prime$. The
primed and the un-primed coordinates are related by the conventional form
of Lorentz transformations and are given by
\begin{eqnarray}
x^\prime&=&\gamma(x-vt), ~~y^\prime=y,~~ z^\prime=z ~{\rm{and}}~ \nonumber \\
t^\prime&=&\gamma\left( t-\frac{vx}{c^2}\right) ~{\rm{with}}~
\gamma=\left (1-\frac{v^2}{c^2}\right )^{-1/2}
\end{eqnarray}
is the Lorentz factor. Next weconsider a uniformly accelerated
frame $S^\prime$. Now for the sake of simplicity, we assume that the 
acceleration of  the test
particle is also along $x$-direction in $S$-frame and is $\alpha_x$. Then in
the primed frame it is given by
\begin{equation}
\alpha_x^\prime=\alpha_x\gamma^{-3}\left(1-\frac{vu_x}{c^2}\right
)^{-3}, ~~{\rm{where}}~~ \alpha_x=\frac{du_x}{dt}
\end{equation}
Therefore for this  particle having constant acceleration in
$S^\prime$-frame, in which it was instantaneously in coincidence with 
its origin at some earlier time, we have
\begin{equation}
\alpha_x^\prime=g~ {\rm{(say)}}~ =\gamma^3\alpha_x
\end{equation}
Here it has further been assumed that the motion of the particle is
also rectilinear and is along $x$-direction with $u_x=v$.
Then with the initial condition that $u_x=0$ at $t=0$, we have
\begin{equation}
u_x=gt\left [1+\left (\frac{gt}{c}\right )^2 \right ]^{-1/2}
\end{equation}
Therefore the proper time of the particle is given by
\begin{equation}
\tau=\frac{c}{g}\sinh^{-1}\left (\frac{gt}{c}\right )
\end{equation}
Hence 
\begin{equation}
ct=\frac{c^2}{g}\sinh\left (\frac{g\tau}{c}\right )
\end{equation}
Then from the expression $u_x=dx/dt$, the velocity of the particle
along $x$-direction and using eqn.(4), we have
\begin{equation}
x=\frac{c^2}{g}\cosh\left (\frac{g\tau}{c}\right )
\end{equation}
Eqns.(6) and (7) are the parametric form of the equation for the
world line of the particle undergoing constant accelerated motion.
Hence the hyperbolic motion of the particle can be represented
by the equation
\begin{equation}
x^2-(ct)^2=\left ( \frac{c^2}{g}\right )^2
\end{equation}
To obtain the Rindler coordinates, which are used for a system
/ observer undergoing constant proper acceleration in a Minkowski
flat space-time, we have followed the scheme described below.
Following \cite{MS,MAX} (see also \cite{BD}),
we have constructed a vector which is 
tangential with
the world line and is proportional to
\begin{equation}
\left (\cosh\left (\frac{g\tau}{c}\right), \sinh
\left (\frac{g\tau}{c} \right )\right )
\end{equation}
These are again the parametric form of coordinates of the points 
moving along 
the proper time axis of the uniformly accelerated observer. Next
consider a vector which is orthogonal to above vector. The
parametric form of which is proportional to
\begin{equation}
 \left (\sinh\left (\frac{g\tau}{c}\right), \cosh
\left (\frac{g\tau}{c} \right )\right )
\end{equation}
Then the parametric form of the world line of a second observer,
which is $h$
distance ahead of the first one as measured by the first observer 
is given by 
\begin{eqnarray}
ct&=&\left (\frac{c^2}{g}+h\right )\sinh \left (\frac{g\tau}{c}\right ) \nonumber 
~~{\rm{and}}~~ \\
x&=&\left (\frac{c^2}{g}+h\right )\cosh \left (\frac{g\tau}{c}\right )
\end{eqnarray}
For the second accelerated observer, the hyperbolic motion is
described by
\begin{equation}
x^2-(ct)^2=\left (\frac{c^2}{g}+h\right )^2=\left
(\frac{c^2}{g}\right)^2 \left (1+\frac{gh}{c^2}\right )^2
\end{equation}
The magnitude of constant acceleration for the second observer is
therefore 
\begin{equation}
g^\prime=g\left (1+\frac{gh}{c^2}\right )^{-1}
\end{equation}
i.e., the two observers undergoing accelerated motions with different
magnitudes of acceleration.

To obtain the Rindler coordinates, we consider an event with space
time coordinate $ct^\prime$ and $x^\prime$ as measured by the first
observer. Further it has been assumed that $x^\prime$ is the distance
from the origin of the first observer and $t^\prime$ is the
corresponding proper-time. Then with respect to an inertial frame, we
have from eqns.(9)-(11) and considering simultaneity of
these events
\begin{eqnarray}
ct&=&\left (\frac{c^2}{g}+x^\prime\right )\sinh\left (\frac{gt^\prime}
{c}\right ) \nonumber  ~~{\rm{and}}~~ \\
x&=&\left (\frac{c^2}{g}+x^\prime\right )\cosh\left (\frac{gt^\prime}
{c}\right ) 
\end{eqnarray}
Hence one can also express the inverse relations
\begin{equation}
ct^\prime=\frac{c^2}{2g}\ln\left (\frac{x+ct}{x-ct}\right )
~~{\rm{and}}~~ x^\prime=(x^2-(ct)^2)^{1/2}-\frac{c^2}{g}
\end{equation}
The Rindler space-time coordinates, given by eqns.(14) and (15) 
are then just an accelerated frame
transformation of the Minkowski metric of special relativity. The
Rindler coordinate transformations change the Minkowski line element from
\begin{eqnarray}
ds^2&=&d(ct)^2-dx^2-dy^2-dz^2  \nonumber ~~{\rm{to}}~~ \\ ds^2&=&\left
(1+\frac{gx^\prime}{c^2}\right)^2d(ct^\prime)^2-{dx^\prime}^2
-{dy^\prime}^2-{dz^\prime}^2
\end{eqnarray}
Since the motion is assumed to be rectilinear and along $x$-direction, 
$dy^\prime=dy$ and $dz^\prime=dz$. The form of the
metric tensor can then be written as
\begin{equation}
g^{\mu\nu}={\rm{diag}}\left (\left (1+\frac{gx}{c^2}\right
)^2,-1,-1,-1\right )
\end{equation}
Now following the concept of dynamics of special theory
of relativity \cite{LL}, the action
integral may be written as (see also \cite{CGH})
\begin{equation}
S=-\alpha_0 \int_a^b ds\equiv \int_a^b Ldt
\end{equation}
Then using eqns.(16) and (18) and putting $\alpha_0=-m_0 c$, where $m_0$ 
is the
rest mass of the particle, the Lagrangian of the particle is given by
\cite{DLM}
\begin{equation}
L=-m_0c^2\left [\left ( 1+\frac{gx}{c^2}\right )^2 -\frac{v^2}{c^2}
\right ]
\end{equation}
where $\vec v$ is the velocity of the particle. The momentum of the
particle is then given by
\begin{equation}
\vec p=m_0\vec v\left [ \left (1+\frac{gx}{c^2} \right )^2
-\frac{v^2}{c^2} \right ]^{-1/2}
\end{equation}
Hence the Hamiltonian of the particle or the single particle energy is 
given by
\begin{equation}
H=\varepsilon_p=m_0c^2 \left (1+\frac{gx}{c^2}\right ) \left (1+
\frac{p^2}{m_0^2c^2}\right )^{1/2}
\end{equation}
In the new sections, using this simple form of
single particle energy, we shall obtain the Saha equations for
photo-ionization process of hydrogen atoms and also for 
electron-positron pair
creation. We shall follow the standard formalism as discussed in the text
books on astrophysics and statistical mechanics \cite{PH,LLS}.
\section{Photo-Ionization of Hydrogen Atoms}
To study the Saha ionization process for hydrogen atoms, we
consider a partially ionized hydrogen plasma, consisting of neutral
hydrogen atoms, hydrogen ions, electrons and electromagnetic
radiation or photons. For the sake of mathematical simplicity, 
the system is assumed to have
cylindrical geometry and the plasma is expanding with a constant
acceleration $g$ along the positive $x$-direction,
which is also the symmetry axis of the cylinder. Then according to the
principle of equivalence, any ionization or de-ionization of the
accelerated particles taking place at some point is equivalent to
their occurrence at rest frame in presence of a gravitational field
$g$. Now it is well known that depending on the amount of 
energy transferred, the interaction
with photon will either excite the hydrogen atoms or ionize them. The
electrons can also be captured by the hydrogen ions, which is the
de-ionization process. Ultimately a dynamic dead lock situation will be 
reached, when the rate of ionization process and the rate of
de-ionization process become just equal, which may be represented by
the reaction equation
\begin{equation}
H_n+\gamma \leftrightarrow H^+ +e^-
\end{equation}
where $H_n$ indicates the $n$th excited state of hydrogen atom with
$n=1$ the ground state. Under such physical condition the
partially ionized hydrogen plasma is assumed to be in thermodynamic 
equilibrium with
all of its constituents. In the equilibrium condition the chemical
potentials of the components are related by the equation
\begin{equation}
\mu(H_n)=\mu(H^+)+\mu(e)
\end{equation}
Since the number of photons is not conserved, $\mu(\gamma)=0$.

Now in the non-relativistic situation, the single particle energy as
shown in eqn.(21) may be approximated as
\begin{equation}
\varepsilon_p=m_0c^2\left ( 1+\frac{gx}{c^2}\right )+ \left (
1+\frac{gx}{c^2} \right ) \frac{p^2}{2m_0}
\end{equation}
Then defining
\begin{equation}
m^\prime=m_0\left (1+\frac{gx}{c^2}\right )^{-1} ~~{\rm{and}}~~ m^{\prime\prime}
=m_0\left( 1+\frac{gx}{c^2}\right )
\end{equation}
we can write
\[
\varepsilon_p=m^{\prime\prime}c^2+\frac{p^2}{2m^\prime}
\]
as the effective single particle energy or Hamiltonian of the
particle.
Now it can very easily be shown from the standard results of text 
books on statistical mechanics \cite{HU,LLS} that 
the number density of a particular component (except for photon)
is given by
\begin{eqnarray}
&n&=\frac{N}{\Delta V} \nonumber \\ &=&\frac{4\pi g_d}{h^3}\int_0^\infty p^2dp
\exp\left [ -\frac{1}{kT}\left ( \frac{p^2}{2m^\prime}+
m^{\prime\prime} c^2-\mu\right ) \right ] \nonumber \\
\end{eqnarray}
where $g_d$ is the degeneracy of the species, $\Delta V=A\Delta x$ is
a small volume element, $\Delta x$ is a small length element in the
$x$-direction at a distance $x$ from the origin
and $A$ is the cross-sectional area of the cylinder.
The length $\Delta x$ is such that $g$ is constant within $\Delta V$.
On evaluating the above integral, we have
\begin{equation}
n=n_Qg_d\exp\left [\frac{1}{kT}\left (\mu- m^{\prime\prime}c^2 \right
) \right ]
\end{equation}
Hence
\begin{equation}
\mu=m^{\prime\prime}c^2 -kT\ln\left (\frac{g_dn_Q}{n}\right )
\end{equation}
This is the general expression for chemical potential for a
particular species in terms of its concentration. Further,
\begin{equation}
n_Q=\left (\frac{2\pi m^\prime kT}{h^2}\right )^{3/2}
\end{equation}
is called the quantum concentration for the particular species in the
mixture. Then using eqn.(23) for chemical
equilibrium condition and eqn.(28), the expression for the chemical
potential,  we have
\begin{equation}
\frac{n(H^+)n(e)}{n(H_n)}=\frac{n_{Q_e}}{g_n} \exp \left (-
\frac{\Delta E_n}{kT}\right ) =R_{g>0} ~{\rm{(say)}}
\end{equation}
where $n(i)$ is the number density for the species $i$ and $g_n=n^2$
is the degeneracy of the neutral hydrogen atom in the $n$th excited
state (since the degeneracy of electrons has been considered
separately in the derivation, the factor $2$ does not appear in the
expression for $g_n$) and
\begin{eqnarray}
\Delta E_n&=&\Delta m^{\prime\prime}c^2= \left ( 1+\frac{gx}{c^2}\right ) 
\Delta mc^2, \nonumber
~~{\rm{with}}~~ \\ \Delta m&=&m(H_n)-m(H^+)-m(e)
\end{eqnarray}
is a measure of excitation energy in the present scenario.

Eqn.(30) gives the ratio of equilibrium concentration at a given
temperature as measured in an accelerated frame or in other wards in
a local rest frame in presence of a constant gravitational field $g$.
Whereas the conventional form of Saha equation, i.e., in an inertial frame
with $g=0$ is given by
\begin{equation}
\frac{n(H^+)n(e)}{n(H_n)}=\frac{{n_{Q_e}}_{g=0}}{g_n} \exp \left (-
\frac{\Delta mc^2}{kT}\right ) =R_{g=0} ~{\rm{(say)}}
\end{equation}
Hence the ratio with $g>0$ and $g=0$ can be written as
\begin{equation}
\frac{R_{g>0}}{R_{g=0}}=\left ( 1+\frac{gx}{c^2}\right )^{-3/2}
\exp\left (-\frac{gx\Delta m}{kT}\right )
\end{equation}
\section{Pair Production at High Temperature}
We next consider the electron-positron pair creation at high
temperature in a local rest frame in presence of a constant
gravitational field $g$. The temperature is assumed to be such 
that $kT$ is larger than at least two times the electron rest mass. 
The pair production process can be represented by the reaction equation
\begin{equation}
\gamma+\gamma \leftrightarrow e^- +e^+
\end{equation}
Similar to the photo-ionization of hydrogen atoms, we assume that the 
interacting electron-positron plasma
along with the photons is in thermodynamic equilibrium. In this case
the chemical equilibrium condition is given by
\begin{equation}
\mu(e^+)+ \mu(e^-)=0
\end{equation}
where $\mu(\gamma)=0$ and
\begin{eqnarray}
\mu(e^-)&=&m^{\prime\prime}(e)c^2-kT\ln \left ( \frac{g_en_Q}{n_{e^-}}\right )
~{\rm{and}}~ \nonumber \\
\mu(e^+)&=&m^{\prime\prime}(e)c^2-kT\ln \left ( \frac{g_en_Q}{n_{e^+}}\right )
\end{eqnarray}
Then from eqn.(35), we have 
\begin{equation}
n_{e^-}n_{e^+}=g_e^2n_Q^2\exp\left( -\frac{2m^{\prime\prime}(e) c^2}{kT}
\right )
\end{equation}
where 
\begin{equation}
n_Q=\left ( \frac{2\pi m^\prime(e) kT}{h^2}\right )^{3/2}
\end{equation}
is the quantum concentration for electron and positron.
Here $m^\prime(e)$ and $m^{\prime\prime}(e)$ are given by eqn.(25)
with $m_0$ replaced by $m(e)$, the electron rest mass.

If the system is assumed to be charge neutral then $n_{e^-}=n_{e^+}$.
Further eqn.(37) can also be expressed as a ratio of the product of
electron-positron concentration with $g>0$ and $g=0$, which is given by
\begin{equation}
\frac{{n_{e^-}n_{e^+}}_{g>0}} {{n_{e^-}n_{e^+}}_{g=0}}
=\left ( 1+\frac{gx}{c^2}\right )^{-3/2} \exp\left ( -\frac{2gxm(e)}
{kT}\right )
\end{equation}
\section{Conclusion}
To analyze the findings of this article, let us consider a set of
uniformly accelerated frames in the space outside the event horizon
of the black hole of mass $M$. Then in accordance with the
principle of equivalence, each of the frames are characterized
by a local acceleration $g$. The position of the frames in Rindler
space are indicated by the spatial coordinates $x_l$, measured
from the centre of the black hole. The values for $g$ and $x_l$ are
obviously different for different uniformly accelerated frames. The
reference frames are distributed uniformly in the space, so that some
of them are at a very close proximity with the event horizon, while
some others are far away from the black hole surface. 

Now consider a
particular frame which is far away from the event horizon, i.e., $x_l
\gg R_s$, where $R_s=2GM/c^2$, the Schwarzschild radius of the black
hole. Then the acceleration $g$ of the frame, or equivalently the
gravitational field of the frame at rest may be expressed in
Newtonian form
$g=GM/x_l^2$. Then it can very easily be shown that the factor as
shown below may be expressed as
\[
1+\frac{gx_l}{c^2}=1+\frac{R_s}{2x_l}
\]
Therefore for $x_l\gg R_s$, one can very easily verify from eqn.(25)
that $m^\prime=m^{\prime\prime}\approx$, the rest mass of the particle.
Further, in such asymptotic condition, it can very easily be shown from
eqns.(33) and (39) that $R_{g>0}/R_{g=0}\approx 1$ and $(n_{e^-}
n_{e^+})_{g>0}/(n_{e^-}n_{e^+})_{g=0} \approx 1$ respectively.

On the other hand, if we consider a frame very close to the surface
of the black hole, or the spatial coordinate 
$x_l$ for this frame is very close to $R_s$, then both the
gravitational field and the temperature can be expressed in blue
shifted forms \cite{TOM} and are given by
\begin{equation}
g(x_l)=\frac{\frac{GM}{x_l^2}}{\left (1-\frac{R_s}{x_l} \right )^{1/2}}
~~{\rm{and}}~~
T(x_l)=\frac{T(\infty)}{\left (1-\frac{R_s}{x_l} \right )^{1/2}}
\end{equation}
Hence we have
\[
1+\frac{gx_l}{c^2}=1+ \frac{\frac{R_s}{2x_l}}{\left ( 1-\frac{R_s}
{x_l}\right )^{1/2}}
\]
Obviously the gravitational field, temperature and the above factor
diverges for $x_l\longrightarrow R_s$. Further, in this limit
$m^\prime \longrightarrow 0$ and $m^{\prime\prime}\longrightarrow
\infty$. It is now a matter of little algebra to show that the
quantum concentration for a particular species in this frame is given
by
\begin{equation}
n_Q=\left (\frac{2\pi kT(\infty)}{h^2}\right )^{3/2} \left [ \left (
1-\frac{R_s}{x_l}\right )^{1/2} +\frac{R_s}{2x_l} \right ]^{-3/2}
\end{equation}
Hence for $x_l\longrightarrow R_s$, we have
\begin{equation}
n_Q=\left (\frac{4\pi kT(\infty)}{h^2}\right )^{3/2} 
\end{equation}
i.e., the blue shifted nature of temperature and gravitational field counter balance each other.

In this case the arguments of the exponential terms for
photo-ionization of hydrogen and for $e^--e^+$ pair creation as
shown by the eqns.(33) and (39) are given by
\[
A1=\frac{R_s}{x_l}\frac{\Delta mc^2}{2kT(\infty)}
~~{\rm{and}}~~
A2=\frac{R_s}{x_l}\frac{m(e)c^2}{2kT(\infty)}
\]
respectively. In these quantities also the divergence of $g$ and
that of $T$ cancel each other in the blue shifted region. 
Then  in the limiting case $x_l\longrightarrow R_s$, we have
\[
A1=\frac{\Delta mc^2}{2kT(\infty)}
~~{\rm{and}}~~
A2=\frac{m(e)c^2}{2kT(\infty)}
\]
On the other hand, in such extreme condition
the pre-exponential terms in both eqns.(33) and (39) become zero, 
making both the ratios equal to zero.
Hence we may conclude that strong gravitational field suppresses
photo-ionization of hydrogen and also $e^--e^+$ pair creation in
chemical equilibrium in a partially ionized hydrogen plasma or in am
$e^--e^+$ plasma respectively. Since the blue shifted nature of $g$
and $T$ are counter balanced, the exponential parts do not take
important role. Whereas the blue shifted form of $g$ in the
pre-exponential terms in eqns.(33) and (39) makes these two
quantities exactly zero in the limiting condition. Therefore
photo-ionization of hydrogen is possible at the surface of a
main-sequence star or a post main-sequence star or at the surface
of a white dwarf. For these stellar objects the strength
of gravitational field is moderate, whereas for a neutron stars,
a very compact stellar objects with extremely strong surface gravitational field, it is not
likely to have hydrogen ions at the vicinity of its surface.

\end{document}